\documentclass[floatfix,reprint,aps,prc,showpacs,showkeys,eqsecnum,amsmath,amsfonts,amssymb,twocolumn,groupedaddress,superscriptaddress,nofootinbib,noeprint]{revtex4-1}
\usepackage[colorlinks=true,linkcolor=red,citecolor=blue,urlcolor=blue]{hyperref}

\usepackage{amssymb}
\usepackage{epsfig} 
\usepackage{amsmath} 
\usepackage{graphicx}

\usepackage{color}
\usepackage{float}
\usepackage{xcolor}
\usepackage{amssymb}
\usepackage{enumitem}

\usepackage{txfonts}

\newcommand{\sat}[1]{{#1}_{\mathrm{sat}}}
\newcommand{\sym}[1]{{#1}_{\mathrm{sym}}}
\newcommand{\eff}[1]{{#1}_{\mathrm{eff}}}
\newcommand{\mmax}{M_G^{\mathrm{max}}}
\newcommand{\MSun}{M_{\odot}}
\newcommand{\nB}{n_{\mathrm{B}}}

\definecolor{red}{rgb}{0.8,0,0}
\definecolor{violet}{rgb}{0.4,0,0.4}
\definecolor{green}{rgb}{0,0.5,0.0}
\definecolor{navy}{rgb}{0.0,0.0,0.6}
\definecolor{orange}{rgb}{0.8,0.2,0.0}
\usepackage[normalem]{ulem}  

\newcommand{\physrep}{Phys. Rep.}
\newcommand{\apjl}{Astrophys. J. Lett.}
\newcommand{\apjs}{Astrophys. J. Suppl.}
\newcommand{\mnras}{Mon. Not. Roy. Astron. Soc.}
\newcommand{\aap}{Astronomy $\&$ Astrophysics}
\newcommand{\nphysa}{Nuc. Phys. A}

\begin{document} 

\title{Frequencies of $f$- and $p$-oscillation modes in cold and hot compact stars}

\author{Vivek Baruah Thapa}
\email{thapa.1@iitj.ac.in}
\affiliation{National Institute for Physics and Nuclear Engineering (IFIN-HH),
RO-077125 Bucharest, Romania}

\author{Mikhail V. Beznogov}
\email{mikhail.beznogov@nipne.ro}
\affiliation{National Institute for Physics and Nuclear Engineering (IFIN-HH), RO-077125 Bucharest, Romania}

\author{Adriana R. Raduta}
\email{Corresponding author: araduta@nipne.ro}
\affiliation{National Institute for Physics and Nuclear Engineering (IFIN-HH), RO-077125 Bucharest, Romania}

\author{Pratik Thakur}
\email{thakur.16@iitj.ac.in}
\affiliation{Indian Institute of Technology Jodhpur, Jodhpur 342037, India}

\date{\today}
\begin{abstract}
  A large collection of equations of state (EOSs) built within the covariant density functional (CDF) theory of hadronic matter and allowing for density dependent (DD) couplings is employed to study polar $f$- and $p$- oscillations of cold and hot compact stars. Correlations between oscillation frequencies of cold purely nucleonic neutron stars (NSs), their global parameters as well as properties of nuclear matter (NM) are investigated by considering a set of models from [Phys.~Rev.~C 107, 045803 (2023)], where a number of constraints on the saturation properties of NM, pure neutron matter (PNM) and the lower bound of the maximal NS mass were imposed within a Bayesian framework. The roles of finite temperature and exotic particle degrees of freedom, e.g., hyperons, $\Delta$-resonances, anti-kaon condensates or a hadron to quark phase transition, are addressed by employing a family of models publicly available on \textsc{CompOSE} and assuming idealized profiles of temperature or entropy per baryon and charge fraction. We find that finite temperature effects reduce the oscillation frequencies of nucleonic stars while the opposite effect is obtained for stars with exotic particle degrees of freedom. When the $\Gamma$-law is employed to build finite temperature EOSs, errors in estimating oscillation modes frequencies are of the order of 10\% to 30\%, depending on the mass. Throughout this work the Cowling approximation is used.
\end{abstract}
\maketitle

\section{Introduction}
\label{sec:Intro}

In the presence of internal, e.g., starquakes caused by a crust crack or a pulsar glitch, a sudden phase transition, magnetic reconfiguration, or external disturbances, e.g, accretion or tidal forces in close eccentric binary system, compact objects such as neutron stars (NSs) are perturbed. Restoration of equilibrium is achieved by a series of oscillations typically classified upon the restoring force. The study of NSs oscillations along with oscillations in the tail of gamma-ray flare emissions from magnetars and, for non-radial modes, gravitational waves (GWs) emission can contribute to a better understanding of NSs interiors and dense matter equation of state (EOS).

Quasi-normal oscillation (QNO) modes of NSs have been thoroughly studied in Newtonian~\cite{Cowling_MNRAS_1941} as well as general relativity~\cite{Thorne_ApJ_1967,Kokkotas_LRR_1999} frameworks. The $f$-, $p$- and $g$- modes, which exist also in ordinary stars, including the Sun, have enjoyed much interest. $f$- and $p$- modes, which will make the focus of this paper, are driven by pressure. The $f$- (fundamental) mode is a stable mode that exists only for non-radial oscillations. Its frequency is proportional to the average density of the star and depends only weakly on the details of the stellar structure. The $f$-mode eigenfunctions have no nodes inside the star and they grow towards the surface. For NSs the $f$-mode frequencies range between 1 kHz and 3 kHz. The $p$- (pressure) modes exist for both radial and non-radial oscillations. Their number is infinite. Their frequencies depend on the time it takes for the acoustic wave to cross the star; in NSs the frequency of the $p_1$-mode, the lowest order mode, ranges between 4 kHz and 8 kHz.

The possibility to infer NSs masses and radii based on joint measurements of at least two QNO modes and, thus, constrain the EOS has been addressed for the first time in Ref.~\cite{Andersson_PRL_1996}, where oscillation frequencies and damping times have been shown to be linked to global properties of NSs through EOS-independent relations. Refs.~\cite{Andersson_MNRAS_1998,Kokkotas_MNRAS_2001,Tsui_MNRAS_2005,Lau_ApJ_2010} confirm these findings. Refs.~\cite{Andersson_MNRAS_1998,Kokkotas_MNRAS_2001} demonstrate, among others, that $\nu_f$ scales with the average density of the star through $\sqrt{M/R^3}$, while $M \nu_p$, $\tau_f R^4/M^3$ and $M/\tau_p$ scale with the compactness $C=M/R$. Here $\nu_f$, $\nu_p$, $\tau_f$ and $\tau_p$ stand for the frequencies and damping times of $f$- and $p$-modes and $M$ and $R$ denote NS mass and radius, respectively. More recent Refs.~\cite{Tsui_MNRAS_2005,Lau_ApJ_2010} prove that $M \nu_f$ and $M/\tau_f$ can be expressed as second order polynomials in compactness ($C=M/R$) to a high degree of accuracy.

\looseness=-1
A systematic investigation of the role of nuclear saturation parameters on the $f$-mode oscillation frequencies has been performed recently in Ref.~\cite{Jaiswal_Physics_2021} within a covariant density functional (CDF) model with non-linear couplings of the scalar-isoscalar ($\sigma$) and vector-isoscalar ($\omega$) meson fields. Various order coefficients in the Taylor expansion of the energy per nucleon of symmetric matter and symmetry energy as a function of deviation from the saturation density of symmetric matter have been found to have a negligible effect on $\nu_f(M)$ for $0.2 \lesssim M/\MSun \lesssim 2.2$. At variance with this, for fixed values of $M$, $\nu_f$ appeared to be positively correlated with the value of the Dirac effective mass of the nucleon at the saturation density ($\sat{n}$) and, to a lesser extent, with the value of $\sat{n}$ itself.

The interplay between properties of nuclear matter (NM), oscillation frequencies and damping times of $f$- and $p$- modes have been investigated also in Ref.~\cite{Kunjipurayil_PRD_2022}. A large bunch of unified EOS models~\cite{Fortin_PRC_2016} derived within the CDF theory or, alternatively, the non-relativistic mean field theory of NM with Skyrme-like effective interactions were employed. The results indicate that 
i) $\nu_f$ of NSs with $1.2 \leq M/\MSun \leq 1.8$ are strongly correlated with the pressure ($P$) of $\beta$-equilibrated matter with densities in the range $\sat{n} \lesssim \nB \lesssim 2.5 \sat{n}$, larger mass NS being sensitive to values of pressure at larger densities, 
ii) $\tau_f(1.4~\MSun)$ is correlated with $P(2 \sat{n})$,
iii) $\nu_p(1.4~\MSun)$ is correlated with $P(\sat{n})$ as well as with the slope of the symmetry energy,
iv) the value of $\tau_f$ is lower (higher) for more (less) compact stars.
No effect of the NSs composition was identified. Frequencies calculated within the Cowling approximation were found to deviate from those calculated within a full general relativity framework by up to 33\% (for $f$-mode) and 14\% (for $p$-mode).

The CDF model with non-linear couplings used in Ref.~\cite{Jaiswal_Physics_2021} has been also employed in Ref.~\cite{Pradhan_PRC_2021} and Ref.~\cite{Pradhan_PRC_2022} to study the impact of hyperons on $f$-mode oscillations. The Cowling approximation and the linearized general relativistic formalism have been used, respectively. Results of Refs.~\cite{Pradhan_PRC_2021,Pradhan_PRC_2022} indicate that hypernuclear stars have larger $\nu_f$ values with respect to their nucleonic counterparts with equal value of gravitational mass. This modification is straightforward to explain based on the radius reduction upon the appearance of hyperons and the dependence of $\nu_f$ on the average density of the star. The frequency range of cold $\beta$-equilibrated hypernuclear stars computed within the linearized general relativity framework is $1.47~\mathrm{kHz} \leq \nu_f \leq 2.45~\mathrm{kHz}$.

Proto-neutron stars (PNSs) have been found to show the same QNO modes as cold $\beta$-equilibrated NSs~\cite{Ferrari_MNRAS_2003,Burgio_PRD_2011,Sotani_PRD_2016,Camelio_PRD_2017,Sotani_PRD_2019}. However, their frequencies and damping times depend on the complex entropy per baryon and lepton fraction profiles, which get modified as the PNS cools down and deleptonizes. Roughly speaking, after bounce and up until the star becomes a cold catalyzed NS, the frequencies of $f$- and $p$-modes increase by several tens of percents up to the values they reach in NSs. A more attentive investigation of the results, however, show that at very early moments in the post bounce evolution, $\nu_f$ \cite{Ferrari_MNRAS_2003,Camelio_PRD_2017} and $\nu_p$ \cite{Camelio_PRD_2017} decrease in time. The long time behavior indicates that the scaling with average density and compactness of these oscillation modes frequencies persists also in hot stars. The early time behavior nevertheless suggests that, if the gradients of entropy per baryon, temperature, lepton and/or charge fractions are too strong, deviations from the above-mentioned trends occur. The quasi-stationary evolution from the PNS stage to the cold $\beta$-equilibrated NS was investigated in Ref.~\cite{Burgio_PRD_2011} using a sequence of constant profiles of entropy per baryon and charge fraction, but with different values in the core and the outer layers. This allowed the authors to separately assess the role of temperature and composition on QNO modes, their conclusion being that entropy gradients are more important than composition-related effects. Ref.~\cite{Camelio_PRD_2017} has also demonstrated that, in addition to stellar mass, QNO frequencies depend on the EOS model.

The first aim of this work is to investigate correlations between the frequencies of $f$- and $p$- oscillation modes of cold NSs on the one hand and properties of NM and NSs on the other hand. To this end, the fiducial set of EOS models recently derived by two of us in Ref.~\cite{Beznogov_PRC_2023} is used. It was obtained within a Bayesian framework by imposing a set of constraints to a family of EOS models derived employing a simplified density dependent (DD) CDF approach. These constraints correspond to properties of saturated NM, density dependence of pressure and energy per nucleon in pure neutron matter (PNM) and the lower limit on the maximal NS mass. Together with other results in literature, e.g., those of Ref.~\cite{Jaiswal_Physics_2021}, where the values of the coupling constants are adjusted such as the values of various NM parameters are modified one by one, our work contributes to a better understanding of the role the NS EOS plays on oscillation modes.

The second motivation of this study is to investigate the effects of finite temperature and exotic particle degrees of freedom (d.o.f.) on the $f$- and $p$-modes. To this end, $\nu_f$ and $\nu_p$ of purely nucleonic and exotic stars with controlled profiles of entropy per baryon (or temperature) and charge fraction are confronted against each other over the mass range $1 \leq M/\MSun \leq 2$. The exotic admixtures that are alternatively considered are the following: $\Lambda$-hyperon; $\Lambda$, $\Sigma^{-,0,+}$, $\Xi^{-,0}$ hyperons; $\Lambda$, $\Sigma^{-,0,+}$, $\Xi^{-,0}$ hyperons and $\Delta^{-,0,+,++}$ nucleonic resonances; $\bar K$-condensates; a hadron to quark phase transition.
Similar to the family of EOSs used for the correlation study of cold NSs, all the models considered here belong to the category of DD CDF models. Moreover, they rely on the same nucleonic effective interaction. Together with the idealized thermodynamic conditions the latter aspect is essential for discriminating the effects of thermal and particle composition.


The article is organized as follows.
In Sec.~\ref{sec:EOS} the EOS models employed in this work are catalogued and some of their properties are reviewed. Thermal effects on equilibrium configurations of spherically-symmetric relativistic stars with various d.o.f. are analysed in Sec.~\ref{sec:MR}. The Cowling formalism we adopt to solve for oscillation modes is briefly reviewed in Sec.~\ref{sec:Cowling}. Sec.~\ref{sec:nu_T=0} investigates correlations between $f$- and $p$-modes frequencies of cold NS, selected global parameters of NS and NM parameters. Thermal effects and role of exotic d.o.f. are discussed in Sec.~\ref{sec:nu_T}. The conclusions are drawn in Sec.~\ref{sec:Conclusions}.
Throughout this paper, we use the natural units with $c = \hbar = k_\mathrm{B} = G = 1$.

\section{EOS models}
\label{sec:EOS}

\begin{table*}
	\caption{
		NM properties of the density dependent effective interactions used in this work. Listed are: saturation density ($\sat{n}$) of SNM; energy per nucleon ($\sat{E}$), compression modulus ($\sat{K}$), skewness ($\sat{Q}$) and kurtosis ($\sat{Z}$) of SNM at $\sat{n}$; symmetry energy ($\sym{J}$), its slope ($\sym{L}$), curvature ($\sym{K}$), skewness ($\sym{Q}$) and kurtosis ($\sym{Z}$) at $\sat{n}$; Dirac effective mass of nucleons in SNM at $\sat{n}$ ($\eff{m}$). For DDB$^{*}$ median values and 68\% confidence intervals are provided.
	}
	\label{tab:NMParams}
	\begin{tabular}{cccccccccccccc}
		\hline\noalign{\smallskip}
		\hline\noalign{\smallskip}
		model &  $\sat{n}$ & $\sat{E}$ & $\sat{K}$ & $\sat{Q}$ & $\sat{Z}$ & $\sym{J}$ & $\sym{L}$ & $\sym{K}$ & $\sym{Q}$ & $\sym{Z}$ & $\eff{m}$ & Ref. \\
		& (fm$^{-3}$) & (MeV) & (MeV) & (MeV) & (MeV) & (MeV) & (MeV) & (MeV) & (MeV) & (MeV) & ($m_n$) & \\
		\noalign{\smallskip}\hline\noalign{\smallskip}
		DDB$^{*}$ & $0.154^{+0.0047}_{-0.0048}$ & $-16.1^{+0.2}_{-0.2}$ & $244^{+30}_{-26}$ & $-52.2^{+150}_{-120}$ & $1400^{+380}_{-730}$ & $31^{+0.85}_{-0.82}$
		& $44.6^{+6}_{-5.7}$ & $-105^{+16}_{-16}$ & $821^{+120}_{-130}$ & $-5470^{+1000}_{-1100}$ & 0.662$^{+0.039}_{-0.042}$ &  \cite{Beznogov_PRC_2023} \\
		DD2 &     0.149      & -16.02 & 242.72 & 168.65   & 5232.56 & 31.67 &  55.04 & -93.23 & 598.14 & -5149.17 & 0.563 & \cite{DD2} \\
		\noalign{\smallskip}\hline
		\noalign{\smallskip}\hline
	\end{tabular}
\end{table*} 

\begin{table*}
	\caption{
		List of EOS models used in this work. For each model we provide information on: considered degrees of freedom; maximum gravitational mass of cold $\beta$-equilibrated NS ($\mmax$); radius of canonical $1.4~\MSun$ NS ($R_{1.4}$); radius of a $2.072~\MSun$ NS ($R_{2.072}$); limits of combined tidal deformability $\tilde \Lambda=16\left[\left( M_1+12 M_2\right) M_1^4 \Lambda_1 + \left( M_2+12M_1\right) M_2^4 \Lambda_2 \right]/13 \left(M_1 +M_2\right)^5$ corresponding to the GW170817 event with an estimated total mass $M_T=2.73^{+0.04}_{-0.01}~\MSun$ and a mass ratio range $0.73\leq q=M_2/M_1 \leq 1$. As in Ref.~\cite{Raduta_EPJA_2021, Raduta_EPJA_2022}, values outside the ranges $11.80~{\rm km} \leq R_{1.4} \leq 13.10$ km~\cite{Miller_ApJL_2021}; $11.41~{\rm km} \leq R_{2.072} \leq 13.69~{\rm km}$~\cite{Riley_ApJL_2021}; $110 \leq \tilde \Lambda \leq 800$~\cite{Abbott_PRX_2019} are marked in bold. n.a. (not available) means that quantities could not be calculated or the calculation is not meaningful (data extracted from \textsc{CompOSE} fall into the first category while those corresponding to DDB$^*$ into the second);
		``---'' means that the quantities do not exist. Other notations are: $q$ stands for quarks; $\Lambda$ denotes the $\Lambda$-hyperon; $\Delta$ is the $\Delta(1232)$ resonance; $Y$ generically denotes the $\Lambda$, $\Sigma^{-,0,+}$ and $\Xi^{-,0}$ hyperons; $K$ respectively stands for kaons. For DDB$^{*}$ median values and 68\% confidence intervals are provided.
	}
	\label{tab:NSProp}
	\begin{tabular}{llcccccl}
		\hline\noalign{\smallskip}
		\hline\noalign{\smallskip}
		model & d.o.f & $\mmax$     & $R_{1.4}$ & $R_{2.072}$ & $\tilde \Lambda$ & $\tilde \Lambda$  & Ref.  \\
		&       & ($\MSun$) & (km)    & (km)      & (q=0.73)           & (q=1)                           \\
		\hline\noalign{\smallskip} 
		DDB$^{*}$               & $N$           & $2.15^{+0.14}_{-0.1}$ & $12.6^{+0.34}_{-0.36}$ & n.a. & $607^{+118}_{-105}$ & $587^{+112}_{-100}$ & \cite{Beznogov_PRC_2023} \\
		HS(DD2)                & $N$            & 2.42 &  {\bf 13.2}      & 13.1 & 799  & 758  & \cite{Hempel_NPA_2010} \\
		BHB(DD2Lphi) & $N, \Lambda$   & 2.10 & {\bf 13.2} & 12.2  & 790  & 757  & \cite{Banik_ApJS_2014} \\
		OMHN(DD2Y)             & $N, Y$         & 2.03 & {\bf 13.2} & ---   & 787  & 756  & \cite{Marques_PRC_2017} \\
		R(DD2YDelta)(1.2;1.1;1.0) & $N,Y,\Delta$ & 2.05& 12.3       & ---   & 470  & 434  & \cite{Raduta_EPJA_2022} \\
		MBB(DD2K)              & $N, K^  -$     & 2.19 & {\bf 13.2} & 13.0  & n.a. & n.a. & \cite{Malik_EPJST_2021}\\
		BBKF(DD2F-SF)1.2       & $N, q$         & 2.15 &      12.2  & 11.4  & 501  & 473  & \cite{Bastian_PRD_2021} \\
		BBKF(DD2-SF)1.8        & $N, q$         & 2.06 & {\bf 11.0} & ---   & 218  & 180  & \cite{Bastian_PRD_2021} \\
		\noalign{\smallskip}\hline
	\end{tabular}
\end{table*}

EOS models used in this paper treat the baryonic component within the CDF theory of strongly interacting matter and employ effective interactions with density dependent couplings of mesonic fields to hadrons.

For the analysis of correlations between oscillation frequencies of cold purely nucleonic NSs and parameters of the EOS, we employ the family of models that corresponds to the fiducial case (``run 5'') of Ref.~\cite{Beznogov_PRC_2023}; herewith it will be referred to as DDB$^*$. Individual EOS models in this family, as well as any other family in Ref.~\cite{Beznogov_PRC_2023}, have been derived within a modified version of the simplified DD CDF model proposed in Ref.~\cite{Malik_ApJ_2022}. Similar to the standard DD CDF models like DD2~\cite{DD2} or DDME2~\cite{Lalazissis_PRC_2005}, the model proposed by Malik et al.~\cite{Malik_ApJ_2022} assumes that nucleon-meson couplings depend on density and expresses them in terms of coupling values at saturation. The vector-isovector $\rho$ field is given the same density dependence as in Refs.~\cite{DD2,Lalazissis_PRC_2005}, while simplified density dependencies are postulated for the scalar-isoscalar $\sigma$ and vector-isoscalar $\omega$ fields. The parameter space associated with this model is six dimensional. Isoscalar and isovector channels are governed by four and two parameters, respectively. Posterior distributions of the input parameters of the model as well as posterior distributions of physical quantities, e.g., NM parameters and NS observables are obtained upon posing, in a Bayesian framework, constraints from nuclear physics, ab initio calculations and astrophysical NS observations. Malik et al.~\cite{Malik_ApJ_2022} opted in favor of a minimal number of constraints: four stem from properties of NM and correspond to the saturation density ($\sat{n}$) of symmetric nuclear matter (SNM), energy per nucleon ($\sat{E}$) and compression modulus ($\sat{K}$) of saturated SNM and symmetry energy at saturation ($\sym{J}$); three correspond to the pressure of PNM at the densities 0.08, 0.12 and 0.16 $\mathrm{fm}^{-3}$, as computed by means of $\chi$EFT at N$^3$LO in Ref.~\cite{Hebeler_ApJ_2013} but with a variance twice larger than the one obtained in Ref.~\cite{Hebeler_ApJ_2013}; a lower limit on maximum NS mass of $2~\MSun$. DDB$^*$ adopts for NM and NS the conditions previously used in Ref.~\cite{Malik_ApJ_2022} but makes a different choice in what regards PNM. The difference consists in accounting, in addition to constraints on the pressure, also for constraints on the energy per nucleon; for both quantities the original variance inferred in Ref.~\cite{Hebeler_ApJ_2013} is assumed. The parameters of the marginalized posterior distributions of various NM parameters and selected properties of NSs are provided in Tables \ref{tab:NMParams} and \ref{tab:NSProp}, respectively. In Table~\ref{tab:NSProp} the compliance with astrophysical NS observations is also reported. 

The effects of finite temperature and exotic particle d.o.f. will be investigated considering a collection of CDF models which rely on the DD2~\cite{DD2} nucleon effective interaction. Preference for DD2~\cite{DD2} is due to its ability to reproduce properties of finite nuclei; NM parameters~\cite{Oertel_RMP_2017}; the density dependence of the energy per nucleon in PNM up to $\approx \sat{n}$, as predicted by $\chi$EFT calculations~\cite{Gandolfi_PRC_2012,Hebeler_ApJ_2013}, see Fig.~12 in Ref.~\cite{Fortin_PRC_2016}; compliance with available constraints from NS observations. The latter include:
i) maximum NS masses higher than $\approx 2~\MSun$~\cite{Antoniadis_Science_2013},
ii) combined tidal deformability of the two NSs in the GW170817 event in the range $110 \leq \tilde \Lambda \leq 800$~\cite{Abbott_PRX_2019},
iii) radius of the canonical $1.4~\MSun$ NS is in the range $13.02^{+1.24}_{-1.06}~\mathrm{km}$~\cite{Miller_ApJL_2019},
iv) radius of a $2.072~\MSun$ NS in the range $11.41~\mathrm{km} \leq R_{2.072} \leq 13.69~\mathrm{km}$~\cite{Riley_ApJL_2021}. For values of various NM parameters and selected properties of NS built upon DD2, see Tables~\ref{tab:NMParams} and \ref{tab:NSProp}, respectively.
The exotic blends that are accounted for have been selected such as to cover most of the mixtures discussed in the literature. They are: $\Lambda$-hyperon; $\Lambda$, $\Sigma^{-,0,+}$, $\Xi^{-,0}$ hyperons; $\Lambda$, $\Sigma^{-,0,+}$, $\Xi^{-,0}$ hyperons and $\Delta^{-,0,+,++}$ nucleonic resonances; $\bar K$-condensates; a hadron to quark phase transition.

\begin{figure}
	\centering
	\includegraphics[width=8.0cm, keepaspectratio]{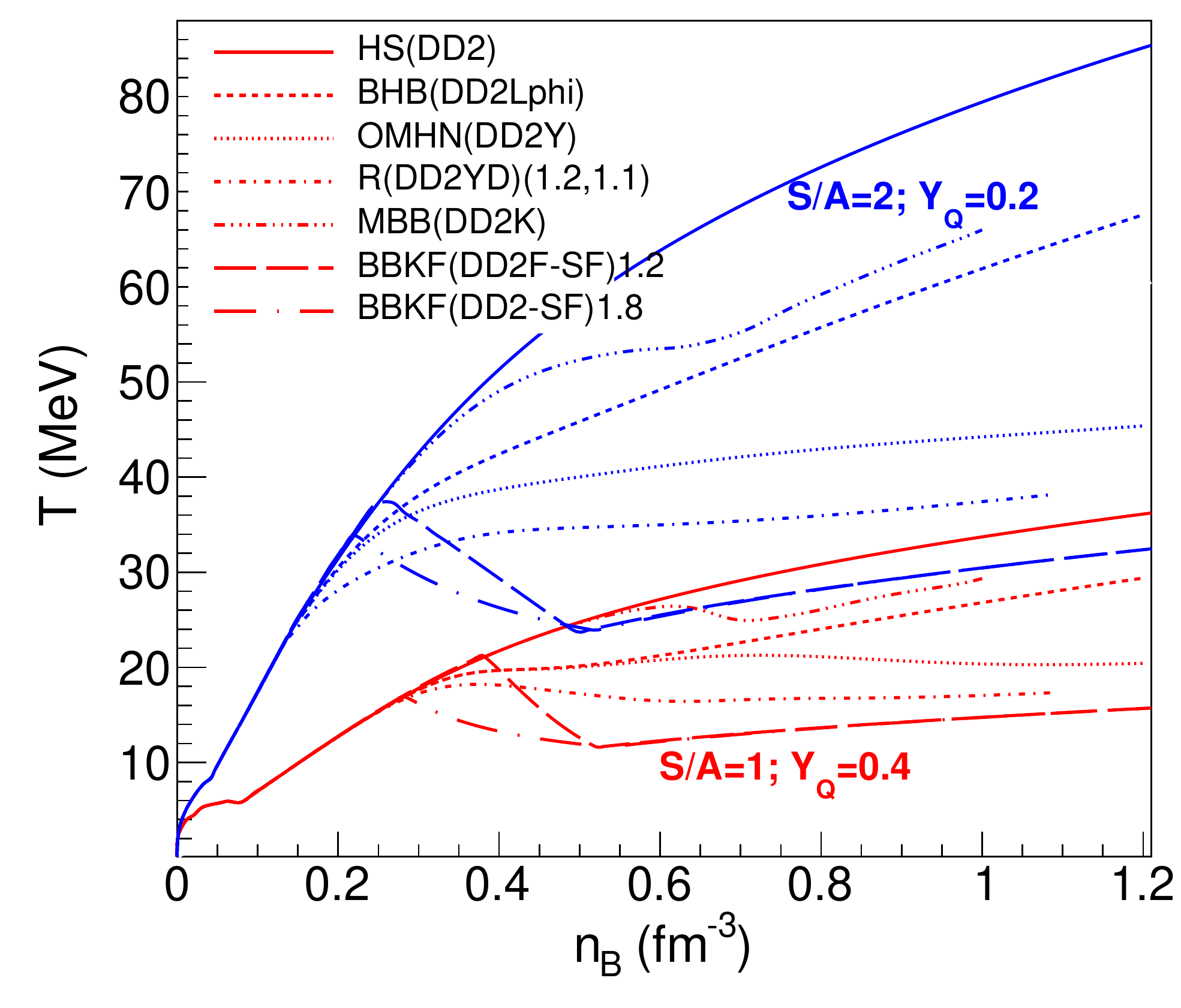}
	\caption{
		Temperature as a function of baryonic number density for ($S/A=1$, $Y_Q=0.4$) and ($S/A=2$, $Y_Q=0.2$). Predictions of EOS models with nucleonic and exotic d.o.f., as mentioned in the legend.
	}
	\label{fig:TnB}
\end{figure}

Three of the models in our set account for strangeness. The model of Ref.~\cite{Banik_ApJS_2014} allows only for $\Lambda$, the less massive hyperon, while the models of Ref.~\cite{Marques_PRC_2017,Raduta_EPJA_2022} account for $\Lambda$, $\Sigma^{-,0,+}$ and $\Xi^{-,0}$. Following standard procedures, in all these models the coupling constants of hyperons to the scalar-isoscalar meson field $\sigma$ are tuned such as to provide for the hyperon at rest in saturated SNM potential well depths in accord with data extracted from hypernuclear experiments~\cite{Gal_RMP_2016}. To be specific, $-30~\mathrm{MeV} \leq U_{\Lambda}^{(N)} \leq -28~\mathrm{MeV}$, $-20~\mathrm{MeV} \leq U_{\Xi}^{(N)} \leq -18~\mathrm{MeV}$ and $U_{\Sigma}^{(N)}=30~\mathrm{MeV}$. Coupling constants of hyperons to vector mesonic fields are fixed using the SU(6) quark flavor symmetry group. The model BHB(DD2Lphi)~\cite{Banik_ApJS_2014} also accounts for $\Lambda \Lambda$ interactions mediated by the hidden vector meson $\phi$; the model OMHN(DD2Y)~\cite{Marques_PRC_2017} also accounts for $YY$ interactions mediated by the scalar $\sigma^*$ and hidden vector $\phi$ mesons.

In addition to the baryonic octet, the model R(DD2YDelta)~\cite{Raduta_EPJA_2022} accounts for $\Delta(1232)$ resonances, which form an isospin quadruplet. The couplings of mesonic fields to $\Delta$s are supposed to have the same density dependence as the couplings to nucleons. The strength of interactions mediated by the $\rho$ meson is considered the same for nucleons and nucleonic resonances, $g_{\rho, \Delta} \left(\nB\right)=g_{\rho, N} \left(\nB\right)$; in what regards the $\omega$ and $\sigma$ mesons, it is assumed that $g_{\omega, \Delta} \left(\nB\right)=1.1 g_{\omega, N} \left(\nB\right)$ and $g_{\sigma, \Delta} \left(\nB\right)=1.2 g_{\sigma, N} \left(\nB\right)$, respectively. The latter choice leads to $U_{\Delta}^{(N)}=-124~\mathrm{MeV}$.

MBB(DD2K)~\cite{Malik_EPJST_2021} accounts for thermal (anti-)kaons and a Bose-Einstein condensate of $K^-$ mesons. The phase transition from the nuclear to antikaon condensed phase is second-order. Nucleons in the antikaon condensed and hadronic phases have different behaviors. Kaon-vector meson couplings are fixed by flavor symmetry arguments. The scalar coupling constant is fixed such that $U_K^{(N)}=-120~\mathrm{MeV}$.

The possibility of a hadron to quark phase transition is addressed in BBKF(DD2(F)-SF)~\cite{Bastian_PRD_2021}. This model assumes that baryonic matter consists of nucleons only and quark matter consists of up and down quarks; quark confinement is modeled within the string-flip model~\cite{Bastian_PRC_2017}. The hadron and quark phases are derived independently and the two phases are then merged through a mixed phase construction. Pure hadron and quark phases are in thermal, mechanical and baryonic chemical equilibrium. The equality of lepton chemical potentials between coexisting phases is replaced, for convenience, by the equality of charge fractions. The two models used here, BBKF(DD2F-SF)1.2 and BBKF(DD2-SF)1.8, differ in the treatment of both baryonic and quark sectors. The nucleon effective interaction DD2F~\cite{DD2F} is a softer version of DD2, designed to recover agreement with data in heavy ion collisions. Its NM properties are identical to those of DD2 and, thus, not listed in Table~\ref{tab:NMParams}.

Properties of cold catalyzed NSs built upon the EOS models listed above are provided in Table~\ref{tab:NSProp}, too. The compliance with observational constraints is also reported. We notice that upon the onset of exotic d.o.f. all the models fulfill the $\approx 2~\MSun$ constraint on the lower bound of maximum NS mass. $\mmax$ larger than $2.072~\MSun$, the estimated mass of the millisecond pulsar PSR J0740+6620~\cite{Fonseca_ApJL_2021}, are obtained only by BHB(DD2Lphi), MBB(DD2K) and BBKF(DD2F-SF)1.2. These models fulfill also the constraint on the radius of massive NS in Ref.~\cite{Riley_ApJL_2021}. The model R(DD2YDelta) provides for $R_{1.4}$ a value by 0.8 km smaller than the one provided by HS(DD2) and, together with BBKF(DD2F-SF)1.2, agrees with constraints from Ref.~\cite{Miller_ApJL_2021}. Models BHB(DD2Lphi), OMHN(DD2Y) and MBB(DD2K) provide for the canonical NS radius the same value as HS(DD2), the reason being that the threshold density for the nucleation of $\Lambda$, the first hyperon to pop up, and $K$ exceed exceed the value of the central density of $1.4~\MSun$ NS. The early transition to deconfined matter in BBKF(DD2-SF)1.8 entails a quite low value for $R_{1.4}$.

Hot stellar matter consists of hadrons or, alternatively, deconfined quarks, leptons and photons. Local densities of strongly interacting particles, e.g., hadrons or quarks, and charged leptons are such that the net charge neutrality condition is fulfilled. Thermal equilibrium as well as chemical equilibrium with respect to the strong interaction are achieved. 

EOSs used in this work, other than those of the DDB$^*$ family, have been extracted from general purpose EOSs tables that are publicly available online on the \textsc{CompOSE}~\cite{COMPOSE_3} site, \url{https://compose.obspm.fr/}. The {\tt compose} software was used.

\begin{figure*}
	\centering
	\includegraphics[width=8.0cm, keepaspectratio]{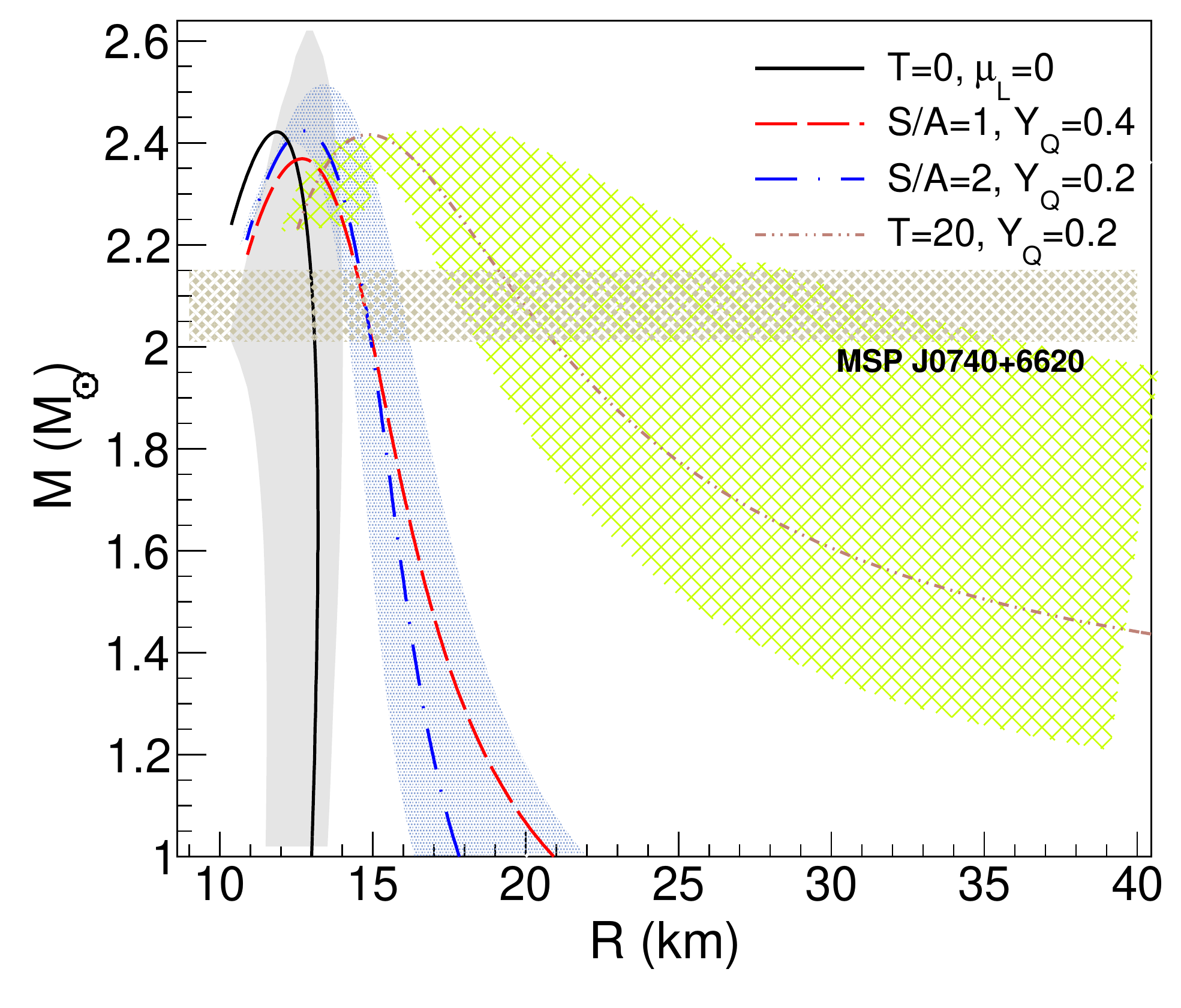}
	\includegraphics[width=8.0cm, keepaspectratio]{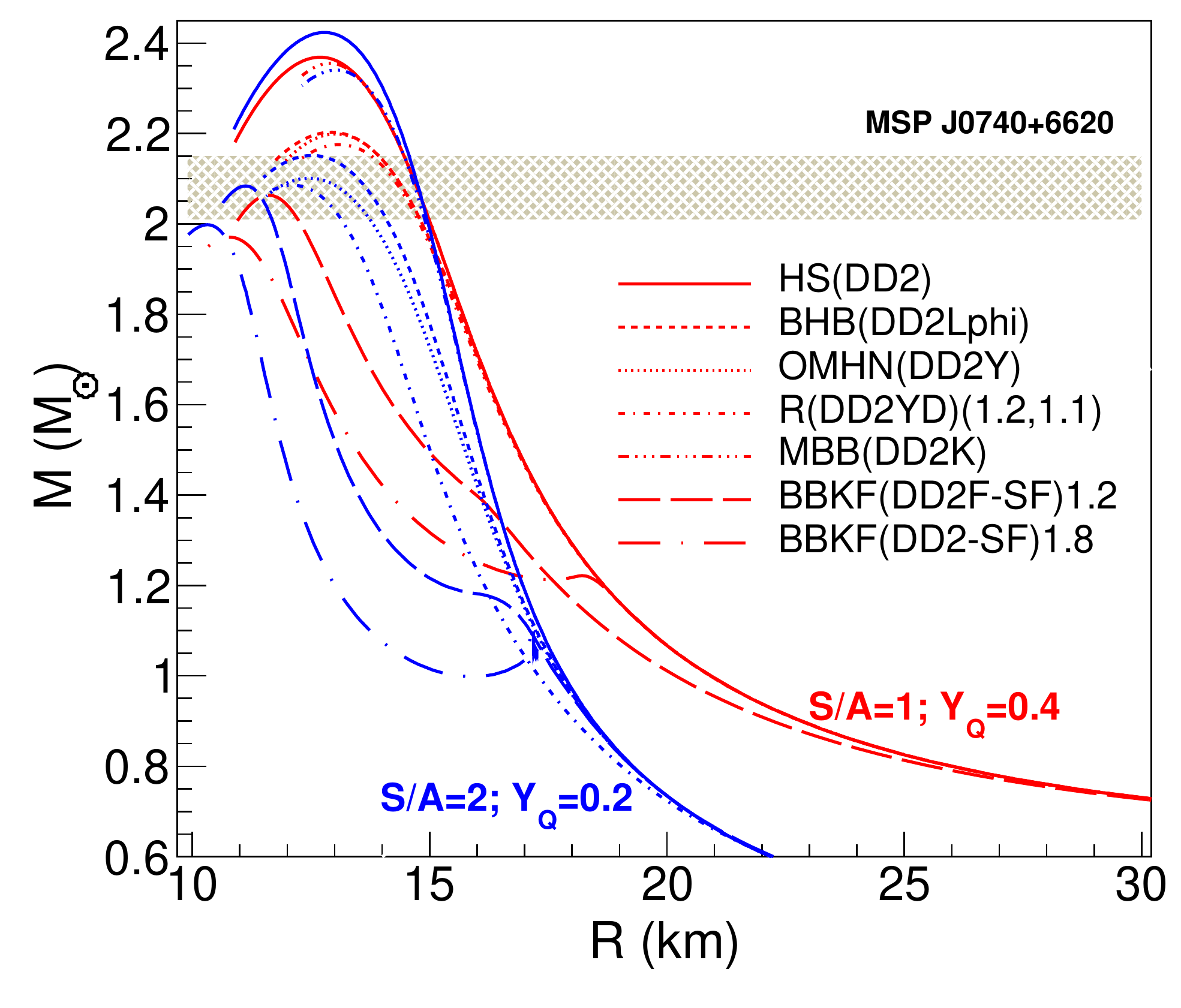}
	\caption{
		Gravitational mass $M$ versus radius for non-rotating spherically-symmetric stars. Various thermodynamic conditions, mentioned on the figures, are considered. Predictions of nucleonic models and models with exotic d.o.f. are illustrated in the left and right panels, respectively. The light green and blue hatched regions in the left panel illustrate the uncertainty domains associated with the use of the $\Gamma$-law with the bounds set to $\Gamma_{\mathrm{th}}=1.5$ and 2; they correspond to ($T=20~\mathrm{MeV}$, $Y_Q=0.2$) and ($S/A=2$, $Y_Q=0.2$), respectively. The light gray shaded domain corresponds to the DDB$^*$ family for cold NSs. The mass constraint from MSP J0740+6620 at $68.3\%$ confidence level~\cite{Fonseca_ApJL_2021} is represented by the horizontal shaded region.
	}
	\label{fig:MR}
\end{figure*}

\section{Equilibrium configurations of compact stars}
\label{sec:MR}

In this section we address the role of finite temperature and exotic d.o.f. on NSs equilibrium
configurations. This is a necessary stage for understanding the modifications that each of these
features brings to oscillation modes.


In order to study in a controlled manner the effect of thermal excitation and the role of various exotic d.o.f., NSs with idealized profiles will be considered in this work. The temperature\,\footnote{In this paper we only speak of local temperatures; redshifted temperatures are not used.} ($T$) profile will be either constant or tuned such as to generate a certain profile for the entropy per baryon ($S/A$). Chemical composition will correspond to fixed charge fractions ($Y_Q$). The thermodynamic conditions considered here are:
i) $S/A=1$, $Y_Q=0.4$, ii) $S/A=2$, $Y_Q=0.2$ and iii) $T$=20~MeV, $Y_Q=0.2$.
Cases i) and ii) correspond to a moment shortly after the core bounce of a core collapsing star and a later time in the evolution from a PNS to a cold deleptonized NS, respectively. Case iii) is purely academic and is considered for pedagogical reasons only.

The temperature profiles for scenarios i) and ii) are plotted as a function of $\nB$ in Fig.~\ref{fig:TnB}. Considered are the DD2-based models of Table~\ref{tab:NSProp}. We note that, in nucleonic matter, the temperature strongly increases with the baryonic density and values as high as several tens MeV are reached. Exotic matter is still hot though definitely less than the nucleonic one. Comparison between temperature values predicted by HS(DD2), BHB(DD2Lphi), OMHN(DD2Y)
and R(DD2YDelta) at $\nB \gtrsim 0.25~\mathrm{fm}^{-3}$ for $S/A=2$, $Y_Q=0.2$ or $\nB \gtrsim 0.6~\mathrm{fm}^{-3}$ for $S/A=1$, $Y_Q=0.4$ indicates that at fixed values of $S/A$ and $\nB$ the temperature drops with the number of particle d.o.f., in agreement with previous results of Ref.~\cite{Raduta_MNRAS_2020}. In some circumstances the drop is strong enough to induce, over a limited density domain, a back-bending of the $T(\nB)$-curve. Examples in this sense are offered by the two BBKF(DD2(F)-SF) models; R(DD2YDelta) and MBB(DD2K) at $S/A=1$, $Y_Q=0.4$. 

\begin{turnpage}
	\begin{figure*}
		\centering
		\includegraphics[]{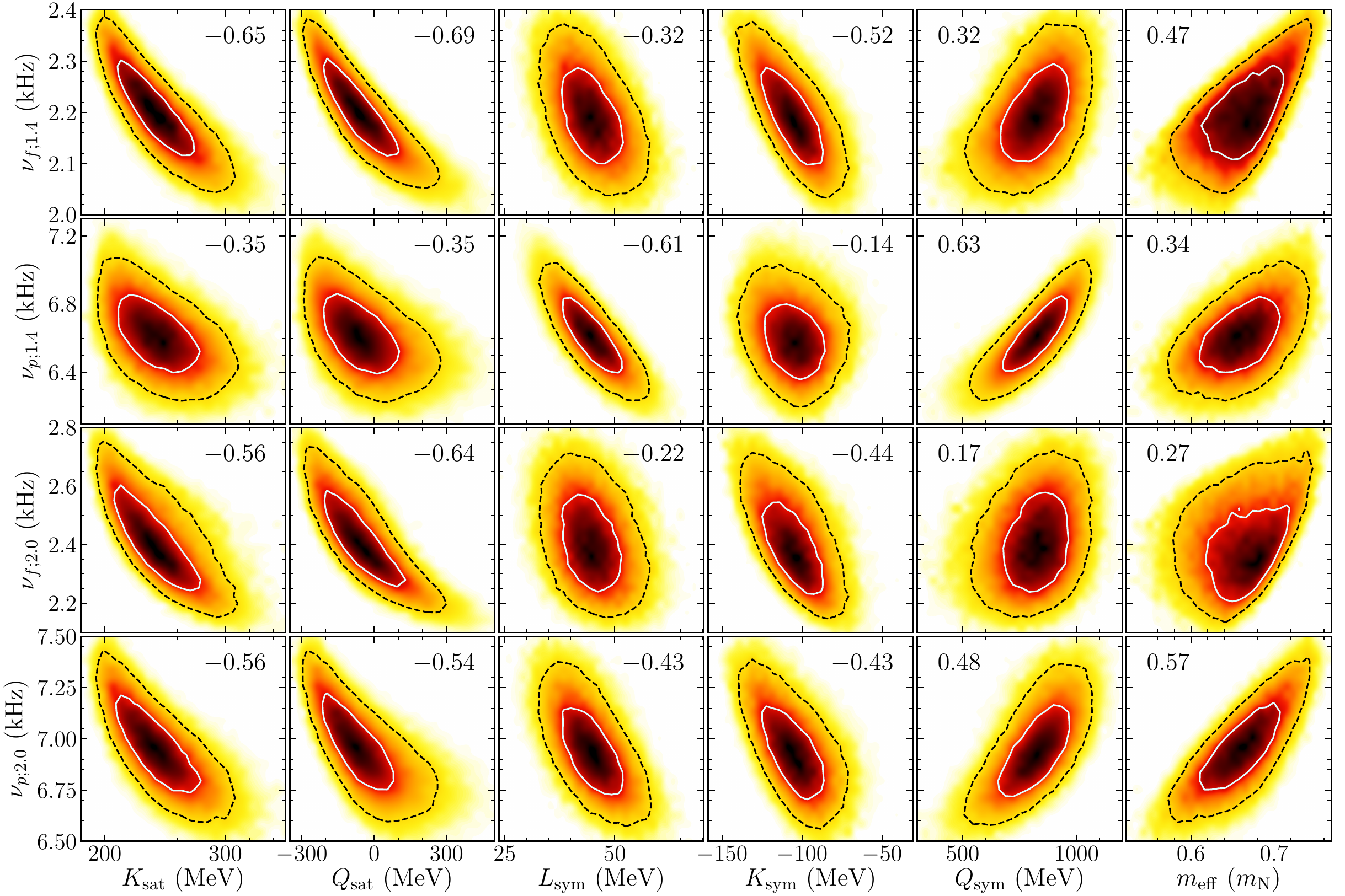}
		\caption{
			Correlations among oscillation frequencies of $f$- and $p$- modes in $1.4~\MSun$ and $2~\MSun$ NSs and selected parameters of NM for the DDB$^*$ family of models. The light cyan solid and black dashed contours demonstrate 50\% and 90\% confidence regions, respectively. The numbers in each panel represent Kendall rank correlation coefficients.
		}
		\label{fig:nu_NM}
	\end{figure*}
\end{turnpage}

The hydrostatic equilibrium of non-rotating spherically-symmetric stars is calculated by solving the Tolman-Oppenheimer-Volkoff (TOV) equations. The $M$-$R$ diagrams of objects with various profiles of $S/A$ (or $T$) and $Y_Q$ (or $\mu_L=0$) and corresponding to different models are depicted in Fig.~\ref{fig:MR}. It comes out that hot stars are more expanded than cold stars and low mass stars are much more affected by finite-$T$ effects than the massive ones. These features were expected and can be explained considering that matter in low mass configurations is more diluted that in massive configurations and dilute matter is more affected by the effects of finite-$T$ than dense matter. We also note that, for $M/\MSun \lesssim 2$, hotter and more isospin asymmetric stars are more compact than the less hot and more isospin symmetric ones. This is in particular the case of ($S/A=2$, $Y_Q=0.2$) versus ($S/A=1$, $Y_Q=0.4$) objects and means that, for densities lower than a ``critical'' value, composition related effects dominate over the temperature related ones. This ``critical'' value obviously depends on particle composition, effective interactions and thermodynamic conditions.
Depending on the d.o.f, the gravitational mass of the most massive configurations augment or diminish with $T$. When the particle composition does not change, as is the case of HS(DD2) and the two BBKF(DD2(F)) models, higher temperatures lead to larger masses. Models where thermally excited species gradually replace the nucleons show the opposite behavior. The latter is the case of models accounting for hyperons, $\Delta$s and $\bar K$-condensates. In what regards the properties of exotic stars, we note that for a given profile of $S/A$:
i) the maximum gravitational mass of exotic stars is smaller than the one of purely nucleonic NS,
ii) for a given value of the gravitational mass, exotic NS are more compact than nucleonic stars,
iii) the larger the number of particle d.o.f. the smaller the maximum mass and the radii of intermediate mass NS.
The same features are observed in cold stars and, as in their case, stem from the EOS softening upon the appearance of exotica. The hadron to quark phase transition in BBKF(DD2-SF)1.8 is responsible for unstable branches and, in the case of $S/A=2$ and $Y_Q=0.2$, also for the occurrence of twin stars. This result is again similar to what is observed in cold NS~\cite{Bastian_PRD_2021}.

Our results concord with those of Refs.~\cite{Prakash_1997,Sumiyoshi_ApJSS_1999,Oertel_EPJA_2016,Marques_PRC_2017,Raduta_MNRAS_2020,Wei_PRC_2021,Khosravi_MNRAS_2022}. Besides, usage of models that employ the same nucleonic effective interaction allows to gauge the role of each extra particle d.o.f.

To assess the reliability of the $\Gamma$-law approximation~\cite{Janka_AA_1993}, we show in Fig.~\ref{fig:MR} also the M-R diagrams obtained when the finite-$T$ EOS is built by employing this recipe. The considered thermodynamic cases are ($S/A=2$, $Y_Q=0.2$) and ($T$=20~MeV, $Y_Q=0.2$).

We remind that the $\Gamma$-law approximation consists in supplementing cold EOS with an ideal gas like component,
\begin{equation}
    \label{eq:Gamma-low}
    P=P_{\mathrm{cold}}+\left(\Gamma_{th}-1 \right) e_{\mathrm{th}},
\end{equation}
where $e_{\mathrm{th}}=e-e_{\mathrm{cold}}$ stands for the thermal energy density and $1.5 \leq \Gamma_{\mathrm{th}} \leq 2$. Similarly, the thermal pressure can be defined as $P_{\mathrm{th}}=P-P_{\mathrm{cold}}$. This approximation was introduced in the '90s as a surrogate for exact finite temperature EOSs~\cite{Janka_AA_1993}, which at that time existed in a very limited number, but it is still in use in numerical simulations~\cite{Bauswein_PRD_2013, Hotokezaka_PRD_2013, Endrizzi_PRD_2018, Camelio_PRD_2019, Huang_PRL_2022}. Its obvious limitation consists in disregarding effects of temperature, density or chemical composition other than those entering $e_{\mathrm{th}}$. For a systematic study of its performances, see Ref.~\cite{Raduta_EPJA_2021,Raduta_EPJA_2022}.

Fig.~\ref{fig:MR} shows that the use of the $\Gamma$-law results in radii uncertainties of the order of 23\%, 13\% and 9\% for models with $M/\MSun=1,~1.4$ and 2 at ($S/A=2$, $Y_Q=0.2$).

\section{Perturbation equations}
\label{sec:Cowling}

In order to solve for nonradial oscillations of spherically symmetric NS the Cowling approximation~\cite{Cowling_MNRAS_1941} is used in this work. It assumes that the spacetime is frozen, which allows one to neglect metric perturbation. The eigenvalues are real, meaning that oscillations are not damped.

The Lagrangian displacement vector of the fluid is given by
\begin{align*}
    \zeta^i=&\left[e^{-\Lambda(r)} W(t,r), -V(t,r) \partial_{\theta}, -V(t,r) \sin^{-2} \theta \partial_{\phi} \right]  \\
    &\times r^{-2} Y_{lm}\left(\theta, \phi \right),
\end{align*}
where $V$ and $W$ are functions of $t$ and $r$ and $Y_{lm}\left(\theta, \phi \right)$ represents the spherical harmonic function. Assuming a harmonic dependence on time, $W(t,r)=W(r) \exp{(i \omega t)}$ and $V(t,r)=V(r) \exp{(i \omega t)}$, mode frequencies are obtained by solving the following system of ordinary differential equations~\cite{Sotani_PRD_2011}:
\begin{align}
	\label{eq:Cowling}
    \frac{dW(r)}{dr}&=\frac{de}{dp} \left[ \omega^2 r^2 e^{\Lambda(r)-2 \Phi(r)} V(r) + \frac{d \Phi(r)}{dr} W(r) \right] 
    \nonumber\\
    &- l \left(l+1\right) e^{\Lambda(r)} V(r), \\
    \frac{dV(r)}{dr}&=2 \frac{d\Phi(r)}{dr} V(r)-\frac{1}{r^2} e^{\Lambda(r)} W(r)\nonumber,
\end{align}
where $\Phi(r)$ and $\Lambda(r)$ are metric functions and $\omega$ stands for the frequency. 

The solution of Eqs.~\eqref{eq:Cowling} with the fixed background metric
\begin{equation}
    ds^2=-e^{2 \Phi(r)} dt^2+e^{2 \Lambda(r)} dr^2 +r^2 d \theta^2 +r^2 \sin^2 \theta d \phi^2,  
\end{equation}
is obtained considering that near the origin $V(r)$ and $W(r)$ behave like
\begin{equation}
  W(r)=A r^{l+1}, ~V(r)=-\frac{A}{l} r^l,
\end{equation}
and the perturbed Lagrangian pressure vanishes on the surface, which leads to
\begin{equation}
    \omega^2 e^{\Lambda(R)-2\Phi(R)} V(R)+\frac 1{R^2} \frac{d \Phi(r)}{dr}|_{r=R} W(R)=0.
    \label{eq:surf}
\end{equation} 

In this work the eigenvalue problem, Eqs.~\eqref{eq:Cowling}, is solved by the shooting method, which consists in generating families of $W(r)$ and $V(r)$ corresponding to various values of $\omega$ and then selecting those which satisfy Eq.~\eqref{eq:surf}. In this paper we shall compute frequencies of the fundamental $f$-mode as well as frequencies of the first $p$-mode with $l=2$. The $f$-mode has no radial nodes, while the first $p$-mode has one radial node.

\section{Correlations between oscillation frequencies of cold NSs, parameters of NM and NS EOS}
\label{sec:nu_T=0}

%
\begin{figure*}
	\centering
	\includegraphics[]{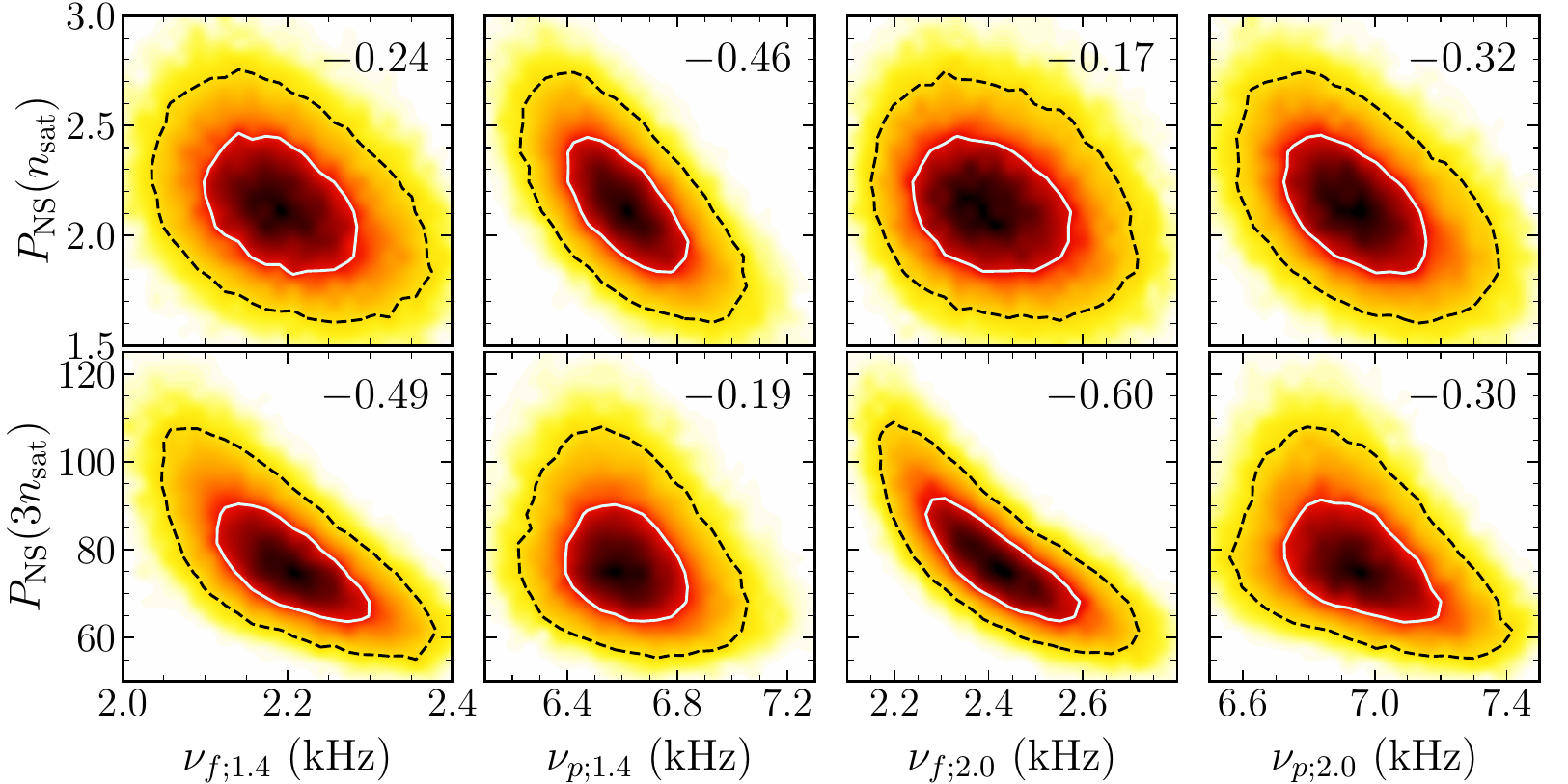}
	\caption{
		Top (bottom) panel: correlations among oscillation frequencies of $f$- and $p$-modes in $1.4~\MSun$ and $2~\MSun$ NSs and the pressure of NS matter at $\sat{n}$ ($3\sat{n}$) for the DDB$^*$ family of models. The light cyan solid and black dashed contours demonstrate 50\% and 90\% confidence regions, respectively. The numbers in each panel represent Kendall rank correlation coefficients.
	}
	\label{fig:nu_pres}
\end{figure*}

Frequencies of $f$- and $p$- oscillation modes have been calculated within the Cowling approximation for the $10^5$ EOS models in the DDB$^*$ family and NSs with masses in the range $1 \leq M/\MSun \leq M_{G}^{\mathrm{max}}$.
Note that each EOS in the DDB* family has its own value of $M_{G}^{\mathrm{max}}$
Then, correlations with parameters of NM, e.g., $\sat{n}$, $\sat{X}$, $\sym{X}$ and $\eff{m}$ and other global properties of NSs or NS EOS have been sought for. Here $X_{\mathrm{sat}}^{(i)}=\left(\partial^i E_0(n_{\mathrm{B}},0)/\partial \mathcal{X}^{(i)}\right)|_{n_{\mathrm{B}}=n_{\mathrm{sat}}}$ and $X_{\mathrm{sym}}^{(j)}=\left(\partial^j E_{\mathrm{sym}}(n_{\mathrm{B}},0) / \partial \mathcal{X}^{(j)}\right)|_{n_{\mathrm{B}}=n_{\mathrm{sat}}}$ denote the parameters of the Taylor expansion of the energy per nucleon of SNM and symmetry energy, respectively, in terms of deviations from saturation $\mathcal{X}=\left(n_{\mathrm{B}}-n_{\mathrm{sat}}\right)/3n_{\mathrm{sat}}$. We start by commenting on the correlations with parameters of NM. Then, we turn to correlations with the pressure of NS matter at densities in the range $\sat{n} \leq \nB \leq 3 \sat{n}$. Finally, we shall investigate correlations with NS radii, average densities and compactness. 

The strongest correlations we have found with parameters of NM are illustrated in Fig.~\ref{fig:nu_NM}, where only NSs with masses $M/\MSun=1.4$ and 2 are considered. On each panel we mention the value of Kendall rank correlation coefficient \cite{Kendall_1938}. We prefer Kendall coefficients to the more commonly used Pearson coefficients due to the non-linearity of the most of our correlations. It comes out that $\nu_f$ is negatively correlated with $\sat{K}$, $\sat{Q}$, $\sym{K}$ and $\nu_p$ is negatively correlated with $\sat{K}$, $\sat{Q}$, $\sym{L}$. These results can be explained considering that $\nu_f$ scales with the average density of the star~\cite{Andersson_PRL_1996,Andersson_MNRAS_1998}; $\nu_p$ scales with NS compactness~\cite{Andersson_MNRAS_1998}; NS radii are positively correlated with $\sat{K}$, $\sat{Q}$, $\sym{L}$ and $\sym{K}$, see Fig.~7 in Ref.~\cite{Beznogov_PRC_2023}. We also note that $\nu_p$ is positively correlated with $\sym{Q}$. This is attributable to the strong and negative correlation between $\sym{Q}$ and $\sym{L}$, see Fig.~3 in Ref.~\cite{Beznogov_PRC_2023}, and to the positive correlation between NS radii and $\sym{L}$ mentioned above. The strength of these correlations obviously depends on NS mass as matter with different densities is probed in light and massive stars. We also note that $\nu_{f;1.4}$ and $\nu_{p;2.0}$ appear to be correlated with $\eff{m}$. Correlations among $\nu_{f}$ and $\eff{m}$ have been previously discussed in Ref.~\cite{Jaiswal_Physics_2021}, where a CDF model with non-linear meson couplings was employed. The other correlation discussed in Ref.~\cite{Jaiswal_Physics_2021}, between $\nu_{f}$ and $\sat{n}$, does not manifest in our model. Two explanations can be envisaged for that. The first trivial one is that this correlation is a peculiarity of the approach used in Ref.~\cite{Jaiswal_Physics_2021}. The second one, that we consider more plausible, is that the limited parameter space exploration allowed by the strongly constrained $\sat{n}$ in Ref.~\cite{Beznogov_PRC_2023} hinders any possible correlation with this quantity.

Fig.~\ref{fig:nu_pres} addresses the correlations between $\nu_f$ and $\nu_p$ and the pressure of NS matter at various densities, previously discussed in Ref.~\cite{Kunjipurayil_PRD_2022}. Our results indicate that $\nu_p$ of low mass NSs is mostly sensitive to the pressure of NS matter at densities around $\sat{n}$ while $\nu_f$ of high mass NSs is mostly sensitive to the pressure of NS matter at densities several times the value of $\sat{n}$. However, none of these correlations are strong.

\begin{figure*}[t]
    \centering
    \includegraphics[]{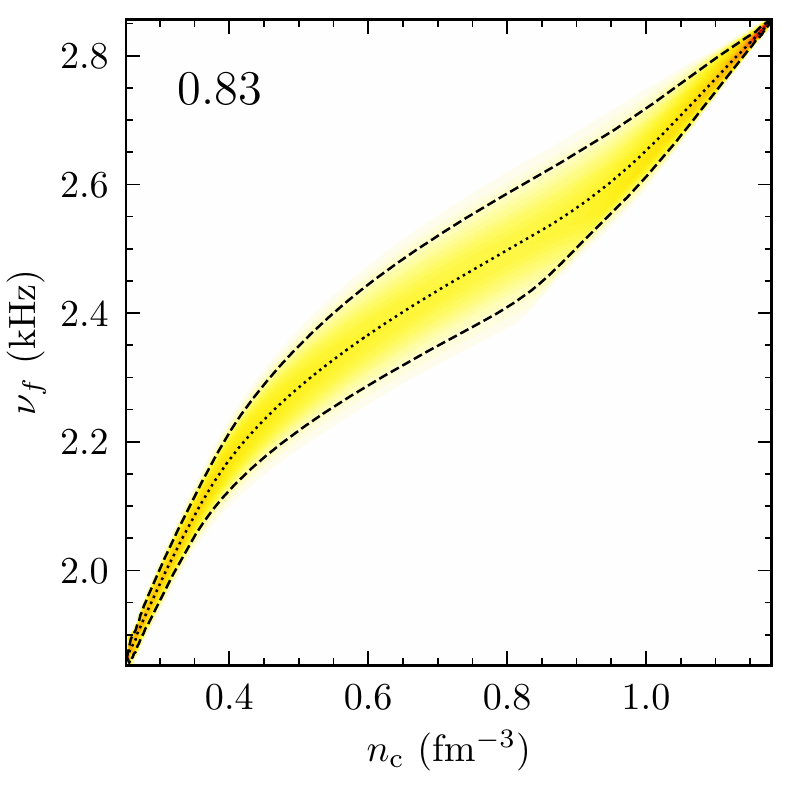}
    \includegraphics[]{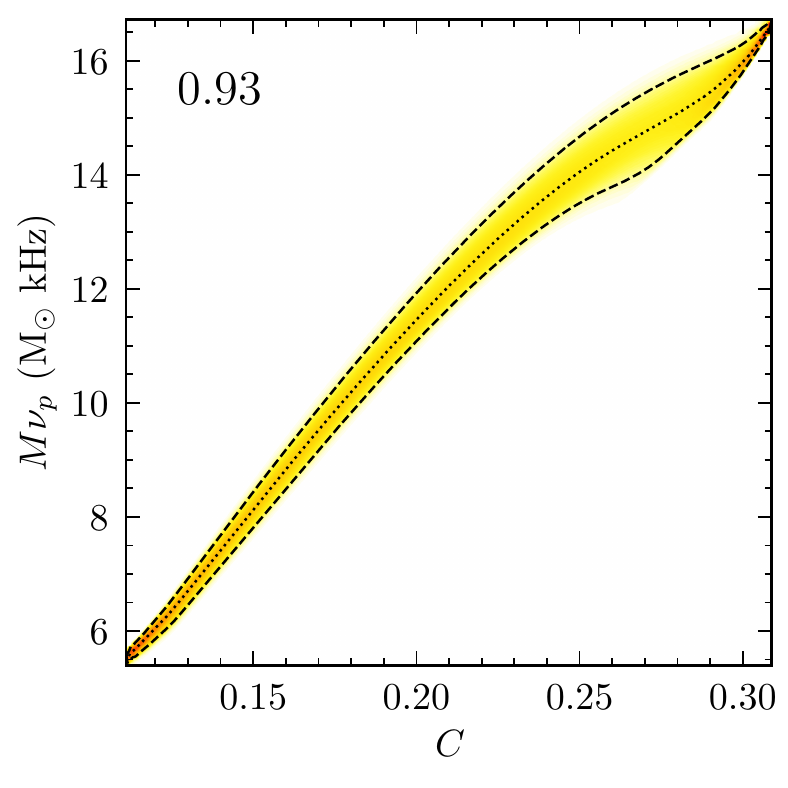}
    \caption{
    Conditional probability density (a.k.a. curves density) plots. Left panel: $f$-mode frequency as a function of central density, $P(\nu_f \,|\, n_\mathrm{c})$. Right panel: the product of $M \nu_p$ as a function of compactness, $P(M \nu_p \,|\, C)$. The results correspond to the DDB$^*$ family of models. The black dashed lines demonstrate 90\% confidence regions. Black dotted lines show to the medians. The numbers in each panel represent Kendall rank correlation coefficients. See text for details.
    }
    \label{fig:scaling}    
\end{figure*}

Fig.~\ref{fig:scaling} shows conditional probability densities (a.k.a. curves densities) corresponding to $\nu_f$, $\nu_p$ and NS properties. The relatively small dispersion of $\nu_f(n_\mathrm{c})$ curves, where $n_{\mathrm{c}}$ represents the central density, suggests that the $f$-mode is mostly sensitive to dense matter properties. We note that, according to our model, this ``correlation'' is of similar strength as the one with the average density, $\sqrt{M/R^3}$, previously put forward in Refs.~\cite{Andersson_PRL_1996,Andersson_MNRAS_1998}. For a quantitative estimate we mention that the values of Kendall rank correlation coefficients are 0.83 and 0.82 for the former and the latter ``correlations'', respectively. An even stronger ``correlation'' links $M \nu_p$ to $M/R$. Indeed, the data corresponding to the $10^5$ models in DDB$^*$ collapse into a relatively narrow band, which suggests that the relation between these two quantities does not depend on the EOS~\cite{Andersson_PRL_1996,Andersson_MNRAS_1998}. One should keep in mind that the term \emph{correlation} is typically used for joint probability density distributions like those plotted on Figs.~\ref{fig:nu_NM} and \ref{fig:nu_pres}. The same holds for correlation coefficients. In this paragraph we use them in a loose sense by applying them to conditional probability density distributions.

Before closing this section let us remind that the bulk of literature devoted to the EOS-dependence of NS properties has shown that the occurrence of correlations as well as their strengths manifest a considerable model dependence. This includes the dependence on the density functional, constraints imposed on the posterior probability density functions and domains of values allowed for both input and output parameters of the model; for a recent discussion, see Ref.~\cite{Beznogov_PRC_2023}. We expect that similar conclusions apply also for correlations involving oscillation frequencies and damping times, though the limited amount of studies available so far can not demonstrate it. As such, it is clear that, in order for a correlation to be considered physical, the conclusions of several different models and approaches have to be confronted. Our present results contribute to this effort.
\begin{figure*}
	\centering
	\includegraphics[width=8.0cm, keepaspectratio=true]{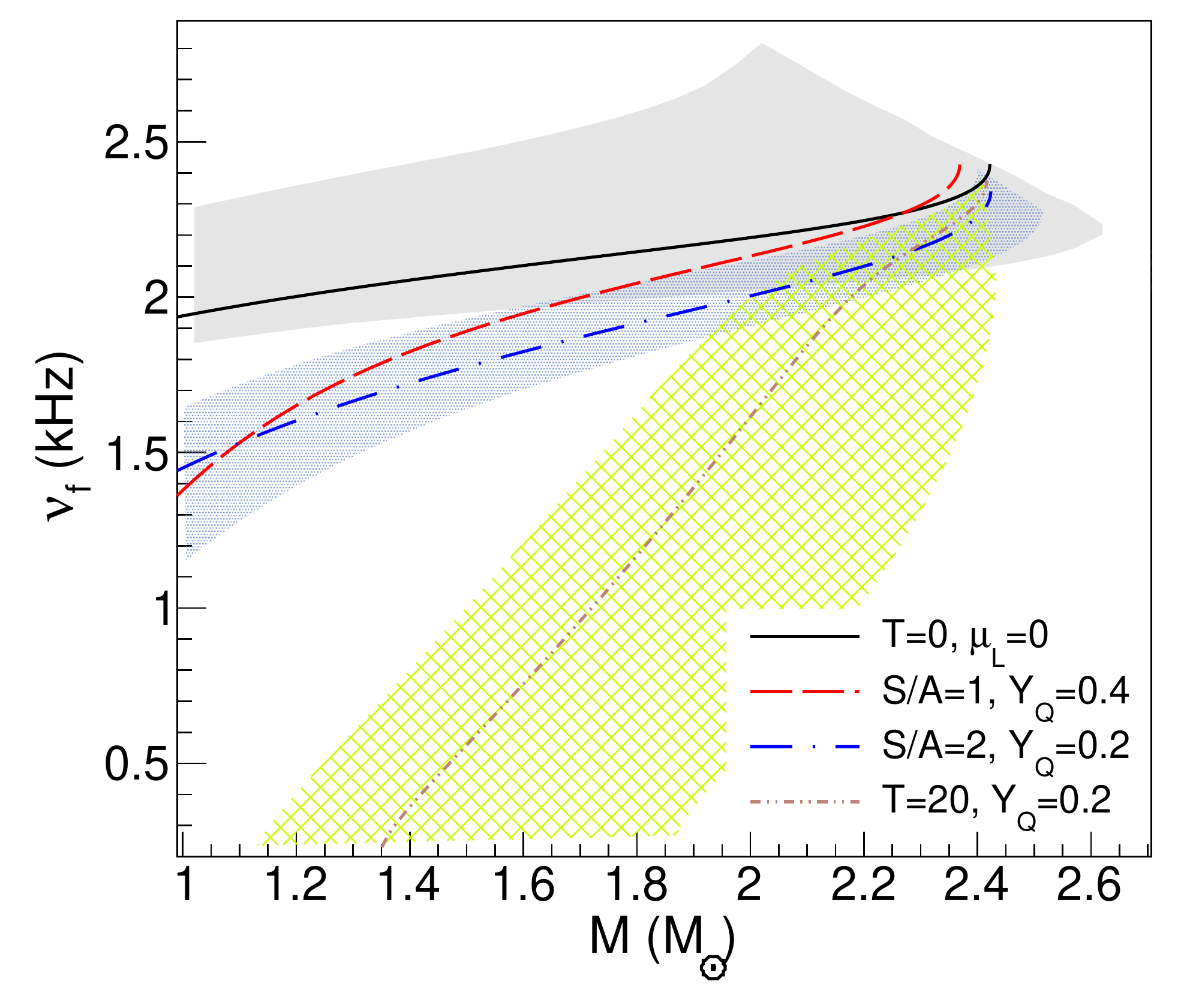}
	\includegraphics[width=8.0cm, keepaspectratio=true]{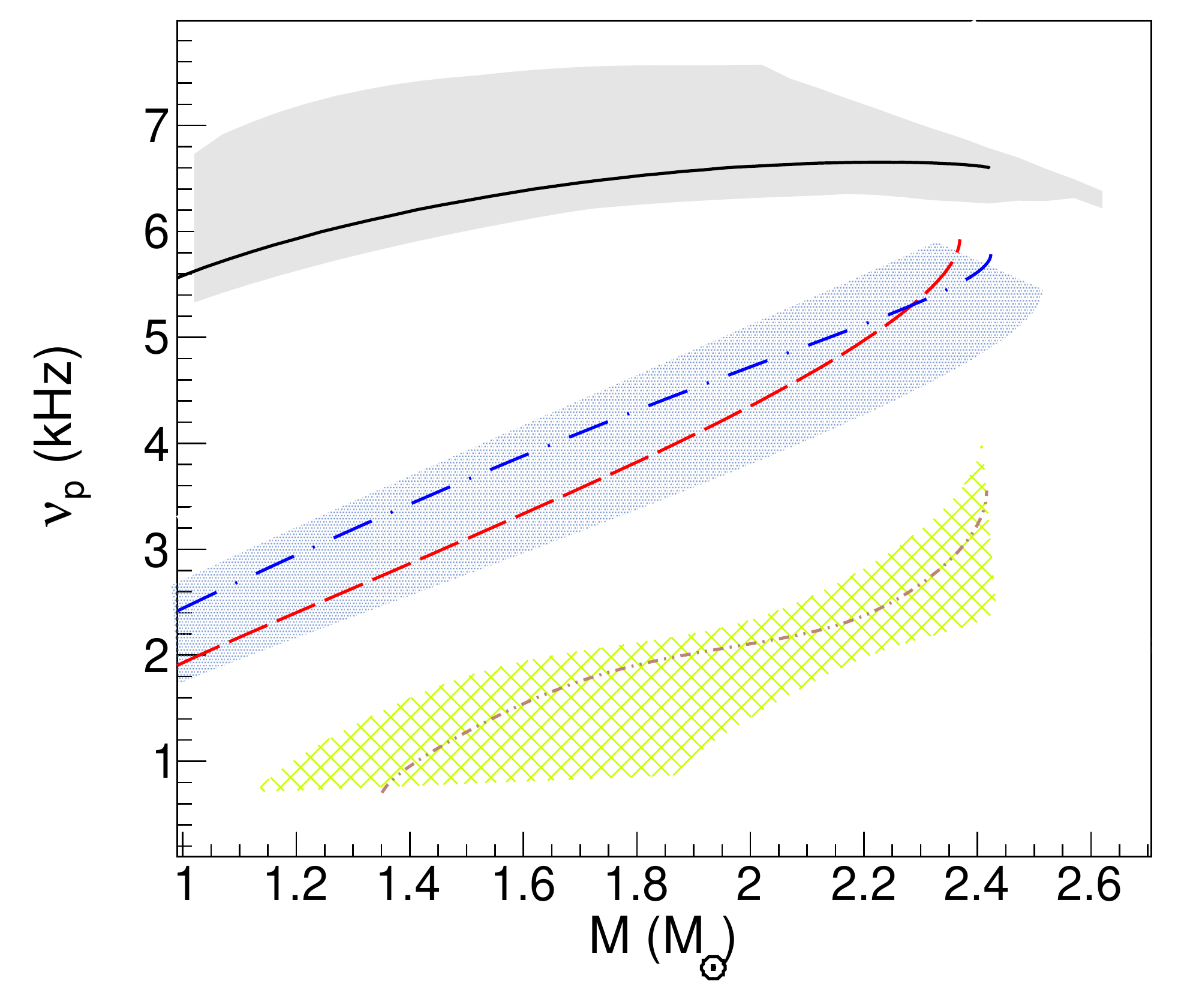}
	\caption{
		$f$- (left panel) and $p$- mode (right panel) frequencies as functions of gravitational mass. The nucleonic models HS(DD2) (curves) and DDB$^*$ (gray shaded regions) are considered. The light green and blue hatched regions illustrate the uncertainty domains associated with the use of the $\Gamma$-law with the bounds set to $\Gamma_{\mathrm{th}}=1.5$ and 2 for ($S/A=2$, $Y_Q=0.2$) and ($T$=20~MeV, $Y_Q=0.2$). Only stable configurations are plotted.
	}
	\label{fig:nu_M_N}
\end{figure*}
%

\section{Thermal and composition effects on oscillation modes}
\label{sec:nu_T}

%
\begin{figure*}
	\centering
	\includegraphics[width=8.0cm, keepaspectratio]{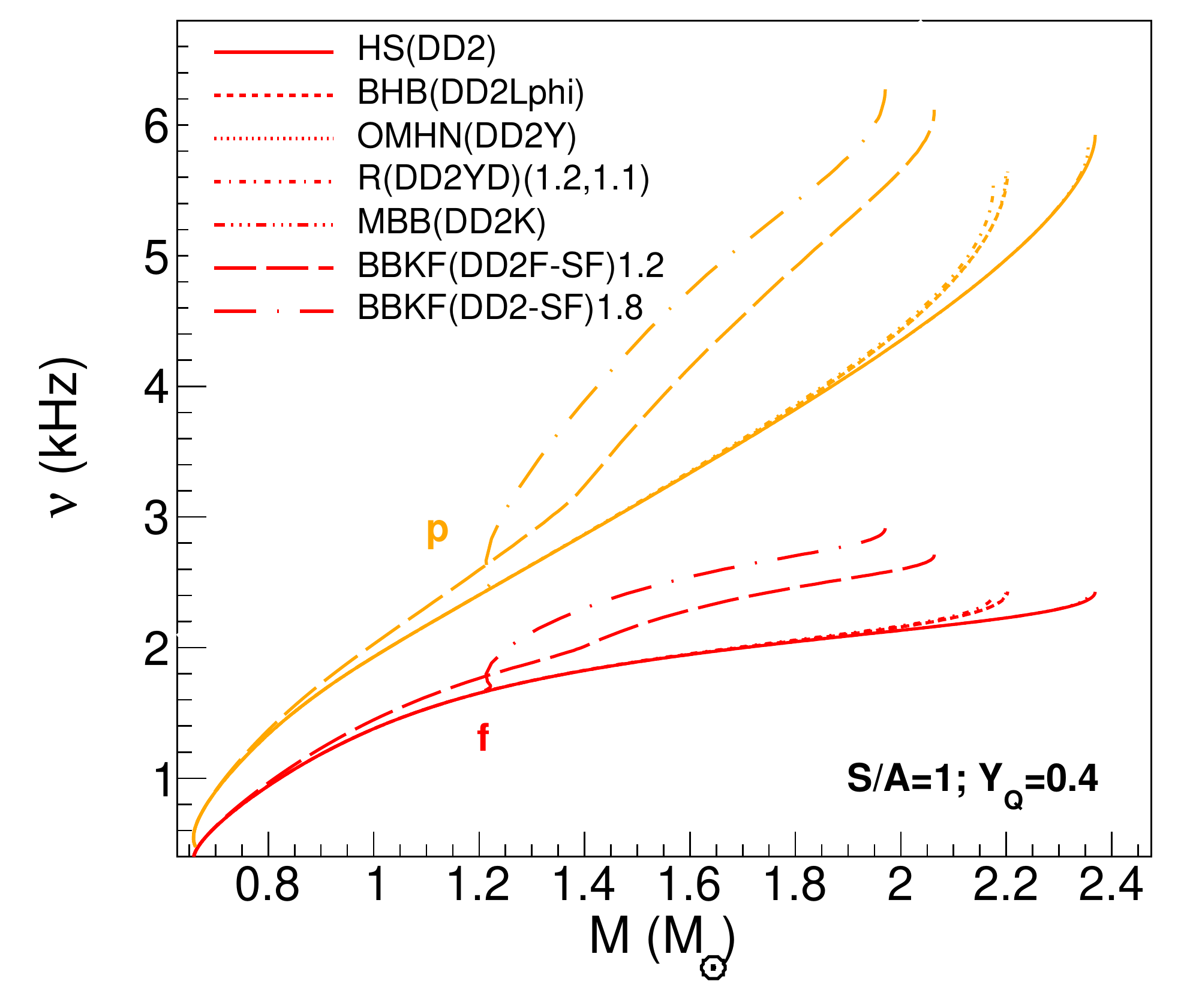}
	\includegraphics[width=8.0cm, keepaspectratio]{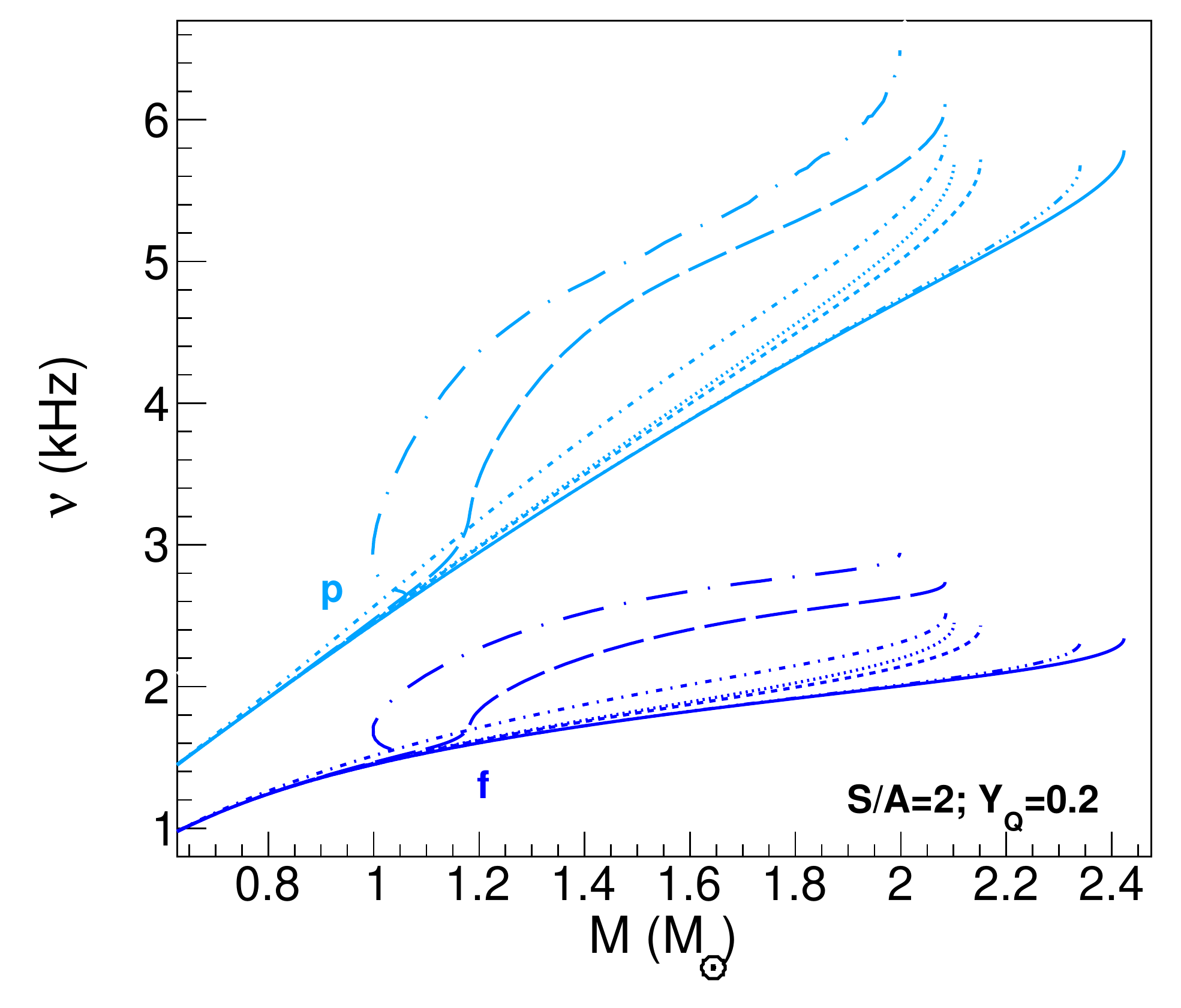}
	\caption{
		$f$- and $p$-mode frequencies as functions of gravitational mass for ($S/A=1$, $Y_Q=0.4$) [left panel] and ($S/A=2$, $Y_Q=0.2$) [right panel]. The considered EOS models are mentioned in the legend. Only stable configurations are plotted.
	}
	\label{fig:nu_M_EXO}
\end{figure*}

Temperature effects on the frequencies of $f$- and $p$-modes of nucleonic stars are investigated in Fig.~\ref{fig:nu_M_N}. The obvious results are that hot NSs have smaller oscillation frequencies than their cold counterparts and the lighter the star the more dramatic this reduction is. The explanation is straightforward considering that both $\nu_f$ and $\nu_p$ scale with a negative power of $R$ and hot stars are more expanded than the cold ones, see Fig.~\ref{fig:MR}. Similar conclusions have been reached in Refs.~\cite{Ferrari_MNRAS_2003,Burgio_PRD_2011}, where more realistic profiles of thermodynamic quantities were considered. We also note that thermal effects influence more the $p$-mode than the $f$-mode. For $f$-mode oscillations, the frequencies corresponding to the maximum mass configurations and different thermodynamic conditions lie close to each other. At variance with this, frequencies of the $p$-mode are scattered. The latter result suggests that
  the two oscillation modes are sensitive to different radial shells in the star.
An extra argument in favor of this assumption is given by the ooposite ranking of $\nu_p$ and $\nu_f$ in stars with $1.2 \lesssim M/\MSun \lesssim 2.1$ at ($S/A=1$, $Y_Q=0.4$) and ($S/A=2$, $Y_Q=0.2$). As for the applicability of the $\Gamma$-law, for models with $M/\MSun=1$, 1.4 and 2 with ($S/A=2$, $Y_Q=0.2$) the use of this approximation results in uncertainties of 22\% (29\%), 9\% (24\%) and 6\% (19\%) for $\nu_f$ ($\nu_p$).

\looseness=-1
Fig.~\ref{fig:nu_M_EXO} shows that models with exotic d.o.f. have higher $\nu_f$ and $\nu_p$ than their nucleonic counterparts. The situation is, again, understandable considering that exotic stars are more compact than nucleonic stars, see Fig.~\ref{fig:MR}, and qualitatively agrees with the results at zero temperature in Ref.~\cite{Pradhan_PRC_2021,Pradhan_PRC_2022}. Out of the models plotted in Fig.~\ref{fig:nu_M_EXO} the most significant increase in frequencies corresponds to R(DD2YDelta) and the two models that account for a hadron to quark transition. These three models experience the most drastic reduction in radii, see Fig.~\ref{fig:MR}. The effect is stronger for ($S/A=2$, $Y_Q=0.2$), where higher values of temperature are reached, than for ($S/A=1$, $Y_Q=0.4$). This means that the modifications entailed by the onset of new d.o.f. dominates over those induced by the temperature and that act in the opposite direction. For stars with $1.6 \leq M/\MSun \leq 2$ at ($S/A=2$, $Y_Q=0.2$) R(DD2YDelta) and BBKF(DD2-SF)1.8 provide values of $\nu_f$ that are by 10\% and 50\% higher than those obtained with HS(DD2). For $\nu_p$ the corresponding figures are 10\% and 35\%.

\section{Conclusions}
\label{sec:Conclusions}

In this paper we have investigated correlations among frequencies of $f$- and $p$- oscillation modes of cold NSs on the one hand and NM parameters, global parameters of NSs and NS EOS on the other hand. The analysis was performed using the $10^5$ models of the DDB$^*$ family that corresponds to the run~5 of Ref.~\cite{Beznogov_PRC_2023}. DDB$^*$ models belong to the DD CDF class and have been obtained in a Bayesian investigation where a certain number of constraints on NM parameters, density dependence of energy per nucleon and pressure in PNM and the lower bound of maximum NS mass have been posed. Our results show that $\nu_f$ is negatively correlated with $\sat{K}$, $\sat{Q}$, $\sym{K}$, the first and latter correlations being stronger in low mass NSs; $\nu_p$ is negatively (positively) correlated with $\sym{L}$ ($\sym{Q}$), both correlations being stronger in low mass stars; massive NSs also manifest correlations among $\nu_p$, $\sat{K}$, $\sat{Q}$; $\eff{m}$ somewhat impacts $\nu_f$ ($\nu_p$) in low (large) mass stars; $P_{NS}(3\sat{n})$ is weakly correlated with $\nu_{f;1.4}$ and more strongly with $\nu_{f;2.0}$; $P_{NS}(\sat{n})$ is weakly correlated with $\nu_{p;1.4}$. Most of these correlations can be explained considering the scaling of $\nu_f$ and $\nu_p$ with NS average density and compactness~\cite{Andersson_PRL_1996,Andersson_MNRAS_1998}. We have also shown that a correlation exists between $\nu_f$ and $n_{\mathrm{c}}$, which suggests that the $f$-mode probes the inner core. Quantitative differences with respect to Ref.~\cite{Jaiswal_Physics_2021,Kunjipurayil_PRD_2022} are illustrative of the model dependence of these results.

The roles of finite-$T$ and exotic particles have been studied by considering a bunch of models that belong to the same DD CDF category, employ the same effective interaction in the nucleonic sector and account for various d.o.f. All these models are available for public use on \textsc{CompOSE}. Two sets of constant profiles of $S/A$ and $Y_Q$ have been adopted to mimic the evolution from a PNS to a cold deleptonized NS. Thermal effects result in a strong reduction of oscillation frequencies in nucleonic stars, they are more important in low mass stars and influence more the $p$-mode than the $f$-mode. Along with the observation that $\nu_f$ of the most massive configurations is only marginally affected by finite-$T$ effects, these indicate that $f$- and $p$-modes
have different sensitivities to different density domains.
Hot stars that allow for hyperons, $\Delta$s, $\bar K$ and quarks have larger values of $\nu_f$ and $\nu_p$ than their nucleonic counterparts. The explanation consists in that the stars with exotica are more compact than the nucleonic stars. Usage of EOSs that rely on the same effective interaction along with consideration of most extra d.o.f. expected to appear at high temperatures and/or densities 
allowed us to comprehend which of these species alter the most the mechanical structure of NSs and whether the pattern changes with the thermodynamic conditions.

Uncertainties associated with the use of the $\Gamma$-law have been gauged by confrontation with the results corresponding to the exact finite-$T$ EOS. In nucleonic stars with constant profiles of $S/A$ they are of the order of several tens of percents and depend sizably on the NS mass and thermodynamic conditions. 

\begin{acknowledgments} 
We acknowledge the funding support from a grant of the Ministry of Research, Innovation and Digitization through Project No. PN-III-P4-ID-PCE-2020-0293. 
\end{acknowledgments}

\bibliography{fp-oscillations}
\end{document}